\input harvmac

\Title{\vbox{\baselineskip12pt
\hbox{BCCUNY-HEP/01-01} \hbox{hep-th/0107092}}}
{\vbox{\centerline{Remarks on Black Hole Degrees of Freedom in
String Theory}}}

\baselineskip=12pt \centerline {Ramzi R.
Khuri\footnote{$^*$}{e-mail: khuri@gursey.baruch.cuny.edu.
Supported by NSF Grant 9900773 and by PSC-CUNY Award 62557 00
31.}}
\medskip
\centerline{\sl Department of Natural Sciences, Baruch College,
CUNY} \centerline{\sl 17 Lexington Avenue, New York, NY 10010
\footnote{$^\dagger$}{Permanent address.}}
\medskip
\centerline{\sl Graduate School and University Center, CUNY}
\centerline{\sl 365 5th Avenue, New York, NY 10036}
\medskip
\centerline{\sl Center for Advanced Mathematical Sciences}
\centerline{\sl American University of Beirut, Beirut, Lebanon
\footnote{$^{**}$}{Associate member.}}

\bigskip
\centerline{\bf Abstract}
\medskip
\baselineskip = 20pt

The Bekenstein-Hawking black hole area entropy law suggests that
the quantum degrees of freedom of black holes may be realized as
projections of quantum states unto the event horizon of the black
hole. In this paper, we provide further evidence for this
interpretation in the context of string theory. In particular, we
argue that increase in the quantum entropy due to the capture of
infalling fundamental strings appears in the form of horizon
degrees of freedom.

\Date{July 2001}

\def\({\left (}
\def\){\right )}
\def\[{\left [}
\def\]{\right ]}

\lref\thorn{See C. B. Thorn, hep-th/9607204 and references
therein; see also O. Bergman and C. B. Thorn, Nucl. Phys. {\bf
B502} (1997) 309.}

\lref\bekhawk {J. Bekenstein, Lett. Nuov. Cimento {\bf 4} (1972)
737; Phys. Rev. {\bf D7} (1973) 2333; Phys. Rev. {\bf D9} (1974)
3292; S. W. Hawking, Nature {\bf 248} (1974) 30; Comm. Math. Phys.
{\bf 43} (1975) 199.}

\lref\GSW {M. B. Green, J. H. Schwarz and E. Witten, {\it
Superstring Theory}, Cambridge University Press, Cambridge
(1987).}

\lref\prep{See M. J. Duff, R. R. Khuri and J. X. Lu, Phys. Rep.
{\bf B259} (1995) 213.}

\lref\cvet{M. Cvetic and D. Youm, Phys. Rev. {\bf D53} (1996)
584; M. Cvetic and A. A. Tseytlin, Phys. Rev. {\bf D53} (1996)
5619.}

\lref\rahm{J. Rahmfeld, Phys. Lett. {\bf B372 } (1996) 198.}

\lref\pol{J. Polchinski hep-th/9611050 and references therein.}

\lref\stva {A. Strominger and C. Vafa, Phys. Lett. {\bf B379}
(1996) 99.}

\lref\malda{J. Maldacena, hep-th/9607235 and references therein.}

\lref\sfet{K. Sfetsos and K. Skenderis, Nucl. Phys. {\bf B517}
(198) 179; R. Argurio. F. Englert and L. Houart, Phys. Lett. {\bf
B426} (1998) 275.}

\lref\holo{G. 't Hooft, gr-qc/9310026; L. Susskind, L. Thorlacius
and J. Uglum, Phys. Rev. {\bf D48} (1993) 3743.}

\lref\corr{L. Susskind, hep-th/9309145.}

\lref\horpol{G. T. Horowitz and J. Polchinski, Phys. Rev. {\bf
D55} (1997) 6189.}

\lref\self {G. T. Horowitz and J. Polchinski, Phys. Rev. {\bf D57}
(1998) 2557.}

\lref\random{P. Salomonson and B. S. Skagerstam, Nucl. Phys. {\bf
B268} (1986) 349; Physica {\bf A158} (1989) 499; D. Mitchell and
N. Turok, Phys. Rev. Lett. {\bf 58} (1987) 1577; Nucl. Phys. {\bf
B294} (1987) 1138.}

\lref\essay{R. R. Khuri, Mod. Phys. Lett. {\bf A13} (1998) 1407.}

\lref\polk{R. R. Khuri, Phys. Lett. {\bf B470} (1999) 73.}

\lref\ent{R. R. Khuri, Nucl. Phys. {\bf B588} (2000) 253.}

\lref\damven{T. Damour and G. Veneziano, hep-th/9907030.}

\lref\liutsey{H. Liu and A. A. Tseytlin, JHEP {\bf 9801} (1998)
010.}

\lref\ferd{R. C. Ferrell and D. M. Eardley, Phys. Rev. Lett. {\bf
59} (1987) 1617.}

\lref\shiraishi{K. Shiraishi, Nucl. Phys. {\bf B402} (1993) 399.}

\lref\mich{D. M. Kaplan and J. Michelson, Phys. Lett. {\bf B410}
(1997) 125; J. Michelson, Phys. Rev. {\bf D57} (1998) 1092.}

\lref\dynam{R. R. Khuri and R. C. Myers, Phys. Rev. {bf D52}
(1995) 6988.}

\lref\gradryz{I. S. Gradshteyn and I. M. Ryzhik, {\it Tables of
Integrals, Series and Products}, (1965) Academic Press (London).}


\newsec{Introduction}

The Bekenstein-Hawking area entropy law \bekhawk\ implies that the
quantum degrees of freedom of black holes may be realized as
projections unto the event horizon of the black hole. While these
degrees of freedom probably ultimately reside in the interior of
the black hole, the holographic principle \holo\ suggests a natural
realization of these degrees of freedom as quantum-gravitational
states on the surface.

String - black hole correspondence \corr\ also has profound
implications for the nature of the underlying quantum degrees of
freedom of black holes. In particular, the fundamental states
underlying black hole entropy, and ultimately quantum gravity from
string theory, arise from string states, even in the strong
coupling limit. While the precise nature of these states in the
black hole limit is far from understood, the fact that not only
the degeneracy but also the precise combinatorics underlying the
counting of constituents are identical in both the quantum string
states picture and the semiclassical black hole entropy
calculation implies that there is perhaps far more to string-black
hole correspondence than what one might reasonably have expected.

In recent work \ent, the implications of correspondence were
considered in two different physical pictures for black hole
formation from strings. The first case concerns the transition,
via an adiabatic increase in the string coupling, from the random
walk picture of free strings \refs{\random,\thorn}\ to a
Schwarzschild black hole at the critical coupling, $g_c$ \self.
The main finding was that the initial, random walk degrees of
freedom transform into holographic, horizon degrees of freedom. In
particular, in both pictures, the entropy of the system is given,
up to a factor of order unity, by an integer representing the
number of random walk steps (or number of string ``bits" in a
polymer string picture), which is also the number of area
``pixels" on the corresponding black hole horizon. This latter
configuration then represents a degeneracy of quantum
gravitational states projected unto the horizon.

In the second case, string - black hole correspondence was applied
to the constituent picture for Reissner-Nordstrom (RN) black hole
solutions in string theory. Here it was pointed out that the
actual combinatorics of counting either string or black hole
constituents was identical in both pictures. This ``constituent
correspondence", in combination with the random walk
correspondence, leads to a particularly simple interpretation of
entropy enhancement in black hole dynamics in correspondence with
analogous processes for string BPS states.

In this paper, we present two calculations in support of
correspondence that show that the underlying degrees of freedom
behind the Bekenstein/Hawking area entropy law \bekhawk\ are
stringlike.  In the first case, we consider a canonical ensemble
of infalling point-like states in a Schwarzschild black hole
background. Using a mean field theory approximation, we show that
the increase in the entropy is accounted for by the number of
steps in the stringy random walk picture. In the second case, we
consider the extremal Reissner-Nordstrom black hole in the
low-velocity approximation. The increase in entropy is again
related in the quantum picture to string degrees of freedom, again
in agreement with the black hole area entropy law.

\newsec{Fundamental String Ensemble in Schwarzschild Background}

It was argued in \refs{\self,\polk}\ that the entropy of a
Schwarzschild black hole in string theory is proportional to $n=
\sqrt{N_s}$, where $N_s$ is the level number of a long-excited
fundamental string which collapses into the Schwarzschild black
hole at the critical value of the string coupling $g_c \sim
N_s^{-1/4} = n ^{-1/2}$. The number $n$ corresponds to the number
of steps (or string ``bits" \thorn\ in the random walk \random\
description of the string at zero coupling. Each step has length
$l_s$, the string scale, with each string ``bit" having mass $m =
m_s = 1/l_s$. The idea is that if we start with this string at
zero coupling and adiabatically increase its string coupling, then
at $g_c$, the string makes a transition into a black hole, such
that the entropies of the string and black hole match up to a
factor of order unity. In the black hole picture, the same number
$n$ represents the number of area ``pixels" in the event horizon.
In this view, the information in the quantum string states are
projected out into the horizon.

Since the arguments behind this picture are rather general, it
would be interesting to see whether this result can be supported
by calculations of the entropy change in a black hole in string
theory as the result of the capture of string states.

To this end, we consider in this section the capture of $n$
fundamental ``test" strings, each of mass $m_s = 1/l_s$, by a
black hole of mass $M = N m_s$, with $1 << n << N$. This black
hole has the same mass as a long, excited fundamental string of
level $N^2$ at zero coupling. This latter string is equivalent to
a random walk with $N$ steps each of length $l_s$.

This ensemble of fundamental strings in $D=4$ may be thought of as
arising, for example, from an ensemble of positively and
negatively electrically charged BPS string states in $D=5$ with
total charge zero. If this latter ensemble collapses to a black
hole, in $D=4$ it will appear neutrally charged, i.e. a
Schwarzschild solution. This is because the effect of the
antisymmetric tensor, $B_{MN}$ and any gauge fields arising from
it due to compactification, are averaged to zero. A similar
analysis to that below was done in \liutsey. The difference
between the two calculations is that in \liutsey, D0-branes of the
same charge are considered, with the consequence of a zero-static
force and even an $O(v^2)$ zero force due to cancellation of
long-range exchange forces. The leading order velocity-dependent
forces are of order $v^4$, but the full Lagrangian is required to
obtain the correct result. In our scenario, we consider a purely
gravitational collapse, as the gauge forces average to zero in the
mean-field limit. In any case, the calculation below for the
Schwarzschild solution is of interest in its own right, whether or
not it arises from a string compactification. Our viewpoint,
however, is that the particle states, each of mass $m$, arise as
pointlike limits of string-like objects with length equal to a
single string scale, $l_s$, corresponding to a single string
``bit" or single step in the random walk picture. The stringy
character of the infalling states, whether point-like or
string-like, is reflected in the random walk picture, with the
corresponding number of extra ``steps" acquired resulting from the
capture of the same number of string bits.

In order to sensibly consider the strings as falling into a black
hole, we must also assume $g^2 N \geq 1$, but $g^2n << 1$, which
is clearly consistent with the above. Assume further that the
coupling is such that the black hole background is not far removed
from an adiabatic formation from a string state: $g > g_c(N)$, so
that the black hole has formed, but $g$ is still of order
$N^{-1/2}$.

The worldsheet action of the test string in a purely gravitational
black hole background is given by \prep\
\eqn\testact{{\cal L}_2 =
-{m\over 2} \sqrt{-\gamma} = -{m\over 2} \sqrt{-\det \gamma_{ij}},}
where the worldsheet metric $\gamma_{ij} =
\partial_i X^M \partial_j X^N g_{MN}$, where $g_{MN}$ is the
background spacetime metric of the black hole. We may equivalently
consider the Lagrangian ${\cal L}_2 = -(m/2) (-\gamma)$, since the
variation of the two actions leads to the the same equations of
motion. In the low-velocity limit, this latter Lagrangian
describes the dynamics of point particles with the same mass as
the string bits. The equivalence of the string and particle
dynamics makes sense provided the Schwarzschild radius $R_s >>
l_s$, which follows from our assumption $n << N$. This assumption
also allows us to neglect the bit-bit interactions.

We now wish to consider the capture of the $n$ particles by the
black hole. For simplicity, assume the entire system is enclosed
in a spherical volume $V = (4\pi/3) R_0^3$. The canonical ensemble
then describes the motion of the particles from $r=R_0$ to the
Schwarzschild radius $R_s = 2 GM$, where we also assume that $R_s
<< R_0$ and where $G$ is the four-dimensional Newton's constant.
Since we would like to mimic an adiabatic process, in which the
total entropy does not undergo a violent change in the process of
the capture, we restrict ourselves to the low-velocity limit.

The results of \ent\ imply that the black hole entropy, whether
seen in the string limit or in the black hole limit, is given by
the total number of random walk steps. So the increase in the
entropy of the black hole as a result of the capture of the $n$
particles is simply $\Delta S \sim n$. This result should be
reproduced from the canonical ensemble calculation of the entropy
of the particles in the black hole background, at least to within
a factor of the order unity.

The Lagrangian for a single particle of mass $m$ in the
Schwarzschild background in the low-velocity limit is given by
\eqn\lagone{{\cal L} = -{m\over 2} \left(\Omega - \Omega^{-1}\dot
r^2 -r^2 \left(\dot\theta^2 + \sin^2\theta
\dot\phi^2\right)\right),}
where $(r,\theta,\phi)$ are spherical
coordinates in three dimensions and $\Omega = 1 -2GM/r$, where $M$
is the mass of the black hole.

In terms of the conjugate momenta, $p_r = \partial {\cal
L}/\partial r$, $p_\theta = \partial {\cal L}/\partial \theta$ and
$p_\phi = \partial {\cal L}/\partial \phi$, the Hamiltonian for
this particle $H = \dot r p_r + \dot \theta p_\theta + \dot \phi
p_\phi - {\cal L}$ is given by
\eqn\hamone{H={m\Omega\over 2} +
{\Omega p_r^2\over 2m} + {p_\theta^2\over 2mr^2} +  {p_\phi^2\over
2mr^2\sin^2\theta}.}
The single particle partition function is
then
\eqn\wone{W=\int d^3 x \int d^3 p \exp{(-\beta H)} = 4\pi
({2\pi m\over \beta})^{3/2} e^{-\beta m\over 2} \int_{2GM}^{R_0}
dr r^2 {e^{\beta G m M\over r}\over \sqrt{\Omega}},}
where $\beta = 1/T_H \sim GM$. Note that the domain of the configuration space
is limited by the horizon of the already formed black hole. Note
also that in our approximation, $\beta m \sim l_p^2 Mm \sim g^2
\l_s^2 Mm \sim g^2 l_s M \sim g^2 N \sim 1$. The integral $I_1$ in
the right hand side of \wone\  may be rewritten as
\eqn\ione{I_1=16 (GM)^3 e^{\beta m\over 2} \int_0^{u_0} {e^{-\beta
m u^2/2}\over (1-u^2)^4} du,} where $u(r)=\sqrt{1-2GM/r}$ and $u_0
= u(R_0)$. The integral $I_2$ in the right hand side of \ione\ can
be estimated as follows:
\eqn\itwo{e^{-\beta m/2} I_3 < I_2 < I_3,}
where $I_3 = \int_0^{u_0} {du\over (1-u^2)^4} \sim
(R_0/R_s)^3$, for $R_s <<R_0$ (see, e.g., \gradryz).
It follows from $\beta m \sim 1$
that $I_2 \sim I_3$ and
\eqn\wtwo{W \sim ({m\over \beta})^{3/2}
(GM)^3 {V\over (GM)^3} = ({m\over \beta})^{3/2} V.}
For $n$ identical (but distinguishable \liutsey) masses, the partition
function is
\eqn\zone{Z=W^n=({m\over \beta})^{3n/2} V^n.}
The energy of the $n$ masses as they approach the horizon is then
given by
\eqn\energy{\delta E=-{\partial \ln Z \over \partial
\beta} = {3n \over 2\beta},}
which is consistent with
equipartition of degrees of freedom, each having energy $(1/2) k_B
T$ (where $k_B$, Boltzmann's constant, has been set to 1
throughout above). The entropy is given by
\eqn\entropy{\delta S = \beta E + \ln Z \sim n.}
This is consistent with the
Bekenstein-Hawking formula, in which
\eqn\entbh{\delta S_{BH} \sim
G M \delta M \sim G M n m \sim n \beta m \sim n,}
where we have used
$\delta E \sim \delta M \sim n m \sim n/\beta$.

So the increase of the black hole entropy can essentially be
accounted for by the addition of the bit degrees of freedom,
arising from a stringy, random-walk picture. This type of
calculation can be repeated for string-like infalling states, with
the same result: the increase in the black hole's entropy is
essentially due to the number of bits acquired.

\newsec{Charged Extremal Black Holes and Constituent Correspondence}

Now we consider the implications of correspondence for extremal
black holes. Here we make use of the ``constituent correspondence"
between string states and singly charged black hole constituents.
For example, a generalization of the extremal Reissner-Nordstrom
black hole solution in four dimensions has metric \cvet
\eqn\rnfour{ds^2
= -\prod_{i=1}^4\left(1+{Q_i\over r}\right)^{-1/2} dt^2 +
\prod_{i=1}^4 \left(1+{Q_i\over r}\right)^{1/2}\left(dr^2 + r^2
d\Omega_2^2\right).}
Following the conventions of \mich, this
solution arises from a compactification from string theory in ten
dimensions  so that $Q_1 = (4G_4 R_9/\alpha') N_1 = N_1 q_{1,0}$,
$Q_2 = (\alpha'/2R_4) N_2 = N_2 q_{2,0}$, $Q_3 = (4G_4/R_9) N_3 =
N_3 q_{3,0}$ and $Q_4 = (R_4/2) N_4 = N_4 q_{4,0}$, where $R_4$ and
$R_9$ are compactification radii for the fourth and ninth spatial
dimensions in $D=10$, respectively. The $N_i >> 1$ represent
eigenvalues of number operators in the string picture (assumed all
to be, say, right movers), with the $q_{i,0}$ representing unit
charges for each of the four species, which therefore have mass
$m_{i,0} = q_{i,0}/l_p^2$.

The area of the black hole is $A = 4\pi R^2 = 4\pi \sqrt{Q_1 Q_2
Q_3 Q_4}$, leading to a black hole entropy $S_{BH} = A/4G_4 = 2\pi
\sqrt{N_1 N_2 N_3 N_4} = S_0$. One of the major recent successes
of string theory has been the recovery of this entropy as a
quantum mechanical entropy due to the the degeneracy of the
charged BPS states corresponding to the $N_i$ \stva. This entropy
can also be interpreted as the number of ``pixels" on the horizon
of the black hole \ent.

For nonextremal black holes, the entropy formula gets generalized
to include number operators for oppositely charged states
\eqn\entgen{S=2\pi (\sqrt{N_1} + \sqrt{\bar N_1}) (\sqrt{N_2} +
\sqrt{\bar N_2}) (\sqrt{N_3} + \sqrt{\bar N_3}) (\sqrt{N_4} +
\sqrt{\bar N_4}).}

As in the above discussion of the Schwarzschild case, we wish to
consider the capture of incoming quanta by the extremal black hole
and attempt to interpret the corresponding increase in entropy in
terms of the quantum degrees of freedom of the system. This
necessarily turns the extremal black hole into a non-extremal one,
even if the incoming charges are all of the right sign, if only
because of the acquired kinetic energy, which results in a
departure from the extremality condition on the total mass (or
energy). For simplicity, consider the capture of $n_1$
constituents/quanta of species ``1", (i.e. each state has charge
$q_{1,0}$), with $1 << n_1 << N_i$. In the limit of zero kinetic
energy (e.g. an adiabatic approach of the quanta), the black hole
simply acquires the extra $n_1$ charges and the extremal entropy
formula is generalized by simply replacing $N_1$ with $N_1 + n_1$
so that $S_1 =S_0 \sqrt{N_1+n_1}/\sqrt{N_1} \simeq S_0 (1 +
n_1/N_1)$, to leading order.

We wish to consider the case in which the kinetic energy of the
incoming quanta, while nonzero, is small. In this case, assume an
average KE of $(1/2) m_{1,0} \bar v^2$, with $\bar v <<1$. Assume
the black hole absorbs these quanta and settles down into a
stationary state. By conservation of energy, the black hole
eventually becomes non-extremal with $N_1 + n_1 + \bar n_1$ right
movers and $\bar n_1$ left movers, such that $\bar n_1 = n_1 \bar
v^2/4 << n_1$. Here the presence of $\bar n_1$ in both left and
right movers is a result of the kinetic energy being transformed
into pairs of right and left movers. The entropy is now given by
\eqn\snew{S_2 = 2\pi (\sqrt{N_1 + n_1 + \bar n_1} + \sqrt{\bar
n_1}) (\sqrt{N_2 N_3 N_4}) = S_1 + 2\pi \sqrt{\bar n_1 N_2 N_3
N_4},} to leading order. The last term represents the gain in
entropy due to the kinetic energy of the quanta. We argue below
that this result is consistent with considerations of
correspondence and heuristic arguments on the quantum degrees of
freedom underlying this entropy.

Unlike the Schwarzschild case, we are unable to use the canonical
ensemble, simply because the temperature of the extremal black
hole is zero. Nevertheless, it is possible from heuristic,
semiclassical considerations to connect the above result to the
possible quantum degrees of freedom of black holes.

The incoming constituents, if stationary, experience zero force
due to the extremal black hole background. This is due to the
cancellation of attractive gravitational and dilatational forces
against repulsive gauge forces as the result of supersymmetry and
the saturation of the BPS bound. When moving in the BH background
in the low-velocity limit, the constituents experience
velocity-dependent forces which, however, lead to {\it velocity
independent} paths, governed by the metric on moduli space. In
other words, in this limit, the path of the quasi-static solutions
are an excellent approximation to the actual paths, being tangent
to them. This is a crucial point in our analysis: the incoming
constituents may arrive in different times, but essentially follow
the same paths in the low-velocity limit. The total energy can
then be divided amongst the different constituents via a simple
partition function.

For the particular case we are considering, the most general
metric on moduli space of this sort was computed in \mich, based
on the work of \refs{\ferd,\shiraishi}. Arguments similar to those
of \dynam\ can also be made to confirm that, by calculating
four-point amplitudes of corresponding string states, it can be
shown that the interaction Lagrangian is velocity-dependent, with
the same form as that of the metric-on-moduli calculation. This
identity is simply the result of supersymmetry.

Without rederiving the results of \mich, it is sufficient to note
that for quanta of species ``1" of mass $m_1$, the leading order
term is a four-body interaction between the four species leading
to the Lagrangian
\eqn\intlag{{\cal L}_{int} = -{m_1\over 2} \left
[ 1 - F(r) \left( \dot r^2 + r^2 )\dot\theta^2 + \sin^2\theta \dot
\phi^2\right) \right],}
where $F(r)=1+{C_{234}\over r^3}$ and
where $C_{234} \sim Q_1 Q_2 Q_3$. In this scenario, the horizon is
initially at $r=0$, but after the capture expands to a radial
position $\delta R$. It is easy to show that $\delta R \sim (Q_1
Q_2 Q_3 Q_4)^{1/4} \sqrt{\bar n_1/N_1} << C_{234}^{1/3}$ provided
we assume that the $N_i$ are of the same order and $n_1 << N_1$.
So in the classical picture, the incoming quanta fall all the way
to $r=0$, but a back-reaction calculation should correct this to
$r \sim \delta R$. In the absence of such a calculation, we
assume, to leading order, that the black hole horizon remains at
$r=0$. So the classical trajectory for the particles in the black
hole background must pass through $r=0$. The quantum (or at least
semiclassical) entropy increase generated by the particles arises
from quantum fluctuations about the classical trajectories.

Since ${\cal L}_{int}$ is independent of $\theta$, angular
momentum is conserved $(1+C_{234}/r^3) r^2 \dot \theta = const. =
L/m$. The radial equation can then be written as
\eqn\radeq{\ddot
r F + {r^2 F'\over 2}={L^2\over m^2r^2 F} ({F'\over 2F} + {1\over
r}).}
This equation can also be obtained from conservation of
total energy
\eqn\conenergy{E = {m\over 2} F(r) (\dot r^2 +
{L^2\over m^2 r^2 F^2}).}
Note that from either \radeq\ or
\conenergy, as $r \to 0$, $\dot r \to 0$, so that the capture does
indeed represent a transformation of the initial kinetic energy of
the particles into additional mass for the black hole. In the
limit $r \to 0$, the radial equation of motion takes the form
\eqn\radzero{\ddot r -{3\dot r^2\over 2r} = -({L^2\over
2m^2 C_{234}^2}) r^3}
with solution $r = \dot r = \ddot r =0$.
Clearly, a more precise analysis would yield a solution with
$r=a$, with $0 < a << C_{234}^{1/3}$ for the above assumptions, as
the horizon must expand with the added mass. Without performing
such a calculation we can still gain some information on the
quantum fluctuations about such a final position for the horizon
from the above equations since, to leading order, they describe
the dynamics of the incoming quanta. Furthermore, the horizon
expansion is a far smaller distance than the length scale of the
equations, so that departures from the correct equations are at
most of order $a$. So now we assume a solution to the corrected
classical equations of the form $r=a$, with  $\dot r = \ddot r =
0$, let us examine the quantum fluctuations about this solution.
Setting $r = a + y$, with $y << a$, and to leading order in $y$,
\radzero\ yields
\eqn\yeqn{\ddot y = - {L^2 a^3\over 2 m^2
C_{234}^2} (1+{3y\over a}).}
In terms of $z = y + a/3$, this
equation has the harmonic oscillator form
\eqn\zeqn{\ddot z = -{3
L^2 a^2\over 2m^2 C_{234}^2}z.}
Note that the shift to the
$z$-variable show that the classical equations are not reliable to
within the order $a$. Nevertheless, the quantum fluctuations can
be understood from \zeqn. This is because the negative sign in the
right hand side of \radzero, coming from \radeq, holds to higher
length scales (or lower order) than $a$, so that whatever the
correct equations including back-reaction may be, the
perturbations about the final horizon position (or capture
position of the particles) will be oscillatory. In other words,
we would expect that a full calculation including the back-reaction
would not affect the quantum oscillations, and hence the quantum entropy.
The significance of this will be discussed below.

Now consider the oscillating modes in \zeqn. These contribute an
energy $\delta E = (n + 1/2)\hbar \omega$ at level $n$, where
$\omega = \sqrt{3L^2 a^2/2m^2 C_{234}^2}$ is the angular frequency
for the oscillations. Setting $\hbar = 1$ and for large numbers,
$\delta E \sim n \omega \sim n La/m C_{234}$. However, $\delta E =
2 m_{1,0} \bar n_1$. Using both expressions for $\delta E$ leads
to $n \sim \bar n_1 c_{234} \sim \bar n_1 N_2 N_3 N_4$. As noted
above, the trajectories of the particles in the low-velocity limit
are independent of the actual velocity, so that an energy level
$n$ can be achieved by any partition of the integer $n$. So the degeneracy
$d(n)$ of the quantum states is given by the partition function
$P(n)$ of the integer $n$. This
implies that the entropy generated by these modes $\delta S = \ln
d(n) \sim \ln P(n) \sim \sqrt{n} \sim \sqrt{\bar n_1 N_2 N_3
N_4}$, in agreement with \snew\ and the area law black hole entropy
above. This argument can easily be generalized for the case of arbitrary
incoming constituents from the four different species: $n_1, n_2, n_3$
and $n_4$. One again finds that the entropy increase $\delta S$ matches
the result expected from the area law. Finally, since the entropy dependence
on the constituents matches in both pictures, one can interpret the area law
entropy in terms of the quantum degrees of freedom available to the constituents
in the process of black hole formation.

Interestingly, if we were to perform the analogous calculation
with the absence of at least one or more of the four species of
constituents, we would obtain equations with no oscillatory modes
at all. For example, if in the above we set $Q_4=0$, then we would
replace $F(r)$ with $G(r) = 1 + C_{23}/r^2$, where $C_{23} \sim
Q_2 Q_3$. Then the negative sign on the right hand side of the
radial equation is replaced with a positive sign, with the
consequence that the equation for quantum fluctuations yields only
decaying modes, instead of oscillatory ones. This leads to no
increase in the entropy, and ultimately to the result that such
black holes possess zero entropy. This is entirely consistent with
the Bekenstein-Hawking expression, since in this case the
``horizon" has zero area and so the black hole entropy is zero.

\newsec{Discussion}

The above arguments are highly heuristic and require more
precise calculations for confirmation. Ultimately, a kind of
``matching" between semiclassical results from general relativity
and quantum degrees of freedom in string theory will lead us to a
firmer understanding of quantum gravity in string theory.
For example, a correct back-reaction calculation already requires
an understanding of this matching, or equivalently, the phase transition
in going from the string to the black hole picture.

The main intent of the above calculations is to further support the notion
of string-black hole correspondence, namely that stringy degrees of freedom are the 
basis for the quantum properties of black holes. It would be interesting
to see whether similar findings can be obtained for charged, non-extremal
black holes, in which combinations of arguments of the previous two sections
can be made. Another possibility is to investigate the case of rotating black
holes and strings with angular momentum.

\listrefs
\end